\documentclass{sig-alternate-05-2015}
\usepackage{paralist}
\usepackage{times,amsmath,epsfig}
\usepackage{amsmath}
\usepackage{amssymb}
\usepackage{stmaryrd}
\usepackage[override]{cmtt}
\usepackage{booktabs}
\usepackage{paralist}
\usepackage{nicefrac}
\usepackage{fancyvrb}
\usepackage{url}
\usepackage{wrapfig}
\usepackage{arydshln}
\usepackage{exscale,relsize}
\usepackage{subfigure}
\usepackage{colortbl}

\newcolumntype{H}{>{\columncolor{black}\color{white}}c}

\graphicspath {{./} {Figures/}}

\begin{document}

\setcopyright{acmcopyright}





%

\title{Big Data Systems Meet Machine Learning Challenges:\\Towards Big Data Science as a Service}
%
%
%
%
%

\numberofauthors{2} 
%
\author{
%
%
\alignauthor
Radwa Elshawi\\
       \affaddr{Princess Nora bint Abdul Rahman University, Saudi Arabia}\\
       \email{rmelshawi@pnu.edu.sa}
\alignauthor
Sherif Sakr\\
       \affaddr{King Saud bin Abdulaziz University for Health Sciences, Saudi Arabia}\\
       \affaddr{University of New South Wales, Australia}\\
       \email{ssakr@cse.unsw.edu.au}
}

\maketitle
\begin{abstract}
Recently, we have been witnessing huge advancements in the scale of data we routinely generate and collect in pretty much everything we do, as well as our ability to exploit modern technologies to process, analyze and understand this data. The intersection of these trends is what is called, nowadays, as \emph{Big Data Science}. Cloud computing represents a practical and cost-effective solution for supporting Big Data storage, processing and for sophisticated analytics applications.
We analyze in details the building blocks of the software stack for supporting big data science as a commodity service for data scientists. We provide various insights about the latest ongoing developments and open challenges in this domain.

\end{abstract}

\section{Big Data Science}
\label{SEC:INTRO}
We live in the \emph{Big Data} era. The continuous growth and integration of data storage, computation, digital devices and networking empowered a rich environment for the explosive growth of big data as well as the tools through which big data is produced, shared, cured  and analyzed~\cite{Sakr16}. The \emph{Big Data} notion  was coined as a  response to the tremendous increase of the world digital data being produced through several means, technologies and in different forms. The notion does not only reflect the size of the data, however, it is usually characterized by the \textbf{3Vs}: 1) \textbf{\emph{V}olume}: refers to the massive amount of data (GBs, TBs, PBs) that is generated and collected. 2) \textbf{\emph{V}elocity}: refers to the  increasing speed and frequency of incoming data that need to be processed. 3) \textbf{\emph{V}ariety}: refers to the diversity on formats (e.g., csv, XML, JSON, PDF), types (e.g., text, images, audio, video), sources and structures (e.g., structured, semi-structured and unstructured) of data from multiple sources.

A McKinsey global report described big data as the next frontier for competition and innovation. The report defined big data as "\emph{Data whose scale, distribution, diversity, and/or timeliness require the use of new technical architectures and analytics to enable insights that unlock the new sources of business value}"~\cite{McKinsey}.  In practice, we are living in an age where a digital revolution coupled with advancements of various emerging technologies including ubiquitous computing devices, sensors and sensing devices, smart devices, cloud computing and big data analytics tools are dramatically changing the mode and accessibility of science, research and practice in all domains.

The Big Data phenomena has urged the scientific communities to reconsider their research methods and processes~\cite{zomaya2017handbook}. In 2007, Jim Gray, the Turing Award
winner, separated data-intensive science from computational science. He called for a paradigm shift in the computing architecture and large scale data processing platforms, Fourth Paradigm~\cite{FourthParadigm}. Experiments, study of theorems and laws, and simulation were in a chronological way the previous three paradigms. Gray argued that this new paradigm does not  only represent a shift in the methods of scientific research, but also a
shift in the way that people think. He declared that the only way to deal with the challenges of this new paradigm is to build a new generation of computing systems to manage, analyze and visualize  the data deluge. Spurred by continuous and dramatic advancements in processing power, memory, storage, and an unprecedented wealth of data, big data processing platforms have been developed in order to tackle the increasingly complex data science jobs. Lead by the \textsf{Hadoop} framework~\cite{HadoopGuide} and its ecosystem, big data processing systems are showing remarkable success in several business and research domains~\cite{Sakr16}.
In particular, for about a decade, the Hadoop platform represented the defacto standard of Big Data analytics world. Though, recently, we have been witnessing a new wave of Big Data 2.0 processing platforms~\cite{Sakr16} that are dedicated to specific verticals such as structured SQL data processing (e.g., \textsf{Hive}~\cite{HIVE1}, \textsf{Impala}~\cite{ImpalaPaper}, \textsf{Presto}\footnote{\url{https://prestodb.io/}}), large scale graph processing (e.g., \textsf{Giraph}~\cite{sakr2016large}, \textsf{Graphlab}~\cite{GraphLab}, \textsf{GraphX}~\cite{GraphX} and large scale stream processing data (e.g., \textsf{Storm}\footnote{\url{http://storm.apache.org/}}, \textsf{Heron}~\cite{kulkarni2015twitter}, \textsf{Flink}~\cite{Flink}, \textsf{Samza}~\cite{noghabi2017samza}, \textsf{Kafka}~\cite{kreps2011kafka}).

In our modern world, data is a basic resource. Although, in principle, data are not useful in and of themselves, they can only be useful if we are able to extract knowledge and value from them. In practice, big data analytics tools enable data scientists to discover correlations and patterns   via analyzing  massive amounts of data from various sources and of different types. Recently, big data science~\cite{DataScience} has emerged as a modern and important  data analysis discipline. It is considered as an amalgamation of classical disciplines such as statistics, artificial intelligence and computer science with its sub-disciplines including database systems, machine learning  and distributed systems. It combines existing approaches with the aim of turning abundantly available data into value for individuals, organizations, and society. The ultimate goal of data science techniques is to convert data into meaningful information.
Both in business and in science, data science methods have shown more robust decision making capabilities.
In practice, in the last few years, we witnessed a huge emergence of big data science in various real-world applications such as business
optimization, financial trading, healthcare data analytics, social network analysis, just to name a few~\cite{Sakr16}.
In particular, we can think of the relationship between big data and data science as being like the relationship between crude oil and an oil refinery.

Continuous developments in various technologies (e.g., sensor computing, mobile computing, ubiquitous computing, cloud computing, social computing) strongly empowered the means to collect and store massive amounts of data in various formats and from different sources. Meanwhile, according to \emph{Moore's Law}, the information density on silicon integrated circuits would double every 18 to 24 months~\cite{schaller1997moore}. In the last decades, we had witnessed  several advances and breakthroughs in our computational capabilities and power. For example, in comparison to a decade before, we currently have
much cheaper, larger, and faster disk storage, memory and computing processor power.
In addition, distributed and parallel computing architectures have enabled the ability for processing massive datasets within reasonable time even with trying exhaustive searches and brute force solutions. Therefore, there are now several big data storage, processing and analytical tools that have made available to turn complex data into meaningful patterns, value and knowledge. Hence, the potential of big data is revolutionizing every aspect of our daily lives.
The techniques and technologies of Big Data Science  have been able to penetrate all facets of the business and research domains.
From the modern business enterprise to the lifestyle choices of today's digital citizen, the insights of big data analytics are driving changes and improvements in every arena~\cite{mayer2013big}. In this article, we analyze in details
the building blocks of the software stack for supporting big data
science as a commodity service for data scientists. In addition, we provide
various insights about the latest ongoing developments and open
challenges in this domain.

\section{Cloud Computing}
Cloud computing represented a paradigm shift for the process of provisioning computing infrastructure. This paradigm shifts the location of this infrastructure to more centralized and larger scale datacenters in order to reduce the costs associated with the management of  software and hardware resources~\cite{ZhaoSLB14}.
It provides its users with the perception of accessing (virtually) unlimited  computing resources where scalability is secured by elastically adding computing resources as the requirement of the workload increases.
It revolutionized the information technology     industry by providing the flexibility to the way that computing resources is consumed by supporting the philosophy of the pay-as-you-go pricing model for the resources and services used by the consumers.
Therefore, cloud computing represented a crucial step towards realizing the long-held dream of envisioning computing as a utility where the economy of scale principles help to effectively drive down the cost of computing infrastructure. In practice, big technology companies (e.g., Amazon, Google, Microsoft,  IBM) have dedicated a lot of resources and investments on establishing their own data centers and cloud-based services across the world to provide assurances on  reliability by providing redundancy for their supported infrastructure, platforms and applications to their cloud consumers.
A recent analysis\footnote{\url{http://research.gigaom.com/2014/11/
big-data-analytics-in-the-cloud-the-enterprise-wants-it-now/}} showed that 53\% of enterprises have deployed (28\%) or plan to deploy (25\%) their Big Data Analytics (BDA) applications in the Cloud.

The original view of cloud computing has been defined by the following three main cloud service models:
\begin{enumerate}
\item \emph{Infrastructure as a Service (IaaS)}: Supports dynamic allocation of the computing resources including data storage, processing,  networks, and other fundamental computing resources to build virtualized systems according to the demands and the requirements of cloud consumers.  Example IaaS providers include \textsf{Amazon Elastic Compute Cloud}\footnote{\url{https://aws.amazon.com/ec2/}}, \textsf{Google Compute Engine}\footnote{\url{https://cloud.google.com/compute/}} and \textsf{Rackspace}\footnote{\url{https://www.rackspace.com/cloud}}.

\item \emph{Platform as a Service (PaaS)}: Supports an abstraction level, which is a software platform on which the system runs. The user does not need to deploy the cloud resources, but has control over the deployed applications and possibly application hosting environment configurations. Example  PaaS providers include \textsf{Microsoft Windows Azure Platform}\footnote{\url{https://azure.microsoft.com/}}, \textsf{Google App Engine}\footnote{\url{https://cloud.google.com/appengine/}}, \textsf{Engine Yard}\footnote{\url{https://www.engineyard.com/}}, \textsf{AppFog}\footnote{\url{https://www.ctl.io/appfog/}}  and \textsf{Heroku}\footnote{\url{https://www.heroku.com/}}.

\item \emph{Software as a Service (SaaS)}: The provider provides services of potential interest to a wide variety of customers hosted in its cloud infrastructure. The services are accessible from various client devices through a thin client interface such as a web browser. The cloud consumers are not required to manage their cloud resources.  Some example services, operating as Software as a Service, that are available in the Internet, include \textsf{Salesforce.com}\footnote{\url{https://www.salesforce.com/}}, \textsf{Oracle}\footnote{\url{http://www.oracle.com/us/products/applications/crmondemand/index.html}}, and \textsf{Zoho}\footnote{\url{https://www.zoho.com/}}.
\end{enumerate}
Fortunately, big data and cloud computing technologies have been combined in a way that has made it easier and more flexible than ever for everyone to step into the world of big data processing. In particular, this technology combination enabled even small companies and individual data scientists  to collect and analyze terabytes of data. For instance, \textsf{Amazon  Elastic Compute Cloud (EC2)} service\footnote{\url{http://aws.amazon.com/ec2/}} is provided as a commodity service which can be purchased and exploited merely by using a credit card to pay for the service. In addition, several cloud-based data storage solution (e.g., \textsf{Amazon Simple Storage Service (S3)}\footnote{\url{https://aws.amazon.com/s3/}}, \textsf{Amazon RDS}\footnote{\url{https://aws.amazon.com/rds/}}, \textsf{Amazon DynamoDB}\footnote{\url{https://aws.amazon.com/dynamodb/}}, \textsf{Google Cloud Data Store}\footnote{https://cloud.google.com/datastore/}, \textsf{Google Cloud SQL}\footnote{\url{https://cloud.google.com/sql/}}), for different data forms, have been provided enabling hosting massive amounts of data at very low cost and on demand. Furthermore, various big data processing frameworks have been made available via cloud-based solutions~\cite{Sakr14}. For  example, Amazon has also released \textsf{Amazon Elastic MapReduce (EMR)}\footnote{\url{http://aws.amazon.com/elasticmapreduce/}} as a cloud service that allows its users to easily and cost-effectively analyze massive sizes of data without the need to get involved in challenging and time-consuming aspects  of running a big data analytics job such as setup, configuration, management, tuning the performance of complex computing clusters. Other cloud-based big data processing services include \textsf{Databricks Spark}\footnote{\url{https://databricks.com/product/databricks}}, \textsf{Amazon Redshift}\footnote{https://aws.amazon.com/redshift/}, \textsf{Google BigQuery}\footnote{\url{https://cloud.google.com/bigquery/}} and \textsf{Azure HDInsight}\footnote{\url{https://azure.microsoft.com/en-us/services/hdinsight/}}.

In practice, these cloud-based services allow third-parties to execute big data analysis tasks over a huge amount of data with minimum effort and cost by abstracting the complexity entailed in developing and maintaining complex computing clusters. Therefore, they paved the way and provided the fundamental elements of the software stack of providing \emph{Big Data Science as a service} in a way that follows the cloud-based trend of providing everything-as-a-service (XaaS)~\cite{banerjee2011everything}.

\section{Big Data Science Frameworks}
The big data phenomena has created ever-increasing pressure for a scalable data processing solution. In addition, the increasing data analysis requirements of almost all application domains have created a crucial requirement for designing and building the new generation of big data science tools that can efficiently and effectively analyze massive amounts of data in order to elicit worthy information, detect interesting
insights and to discover meaningful patterns and knowledge.

\textsf{R}\footnote{\url{https://www.r-project.org/}} is currently considered as the defacto standard in statistical and data analytics research.
It is the most popular open source and cross platform software which has  very wide community support.
It is flexible, extensible and comprehensive for productivity.
\textsf{R} provides a programming language which is used by statisticians and data scientists to conduct data analytics tasks and discover new insights from data by exploiting techniques such as clustering, regression, classification and text analysis. It is equipped with very rich and powerful library of packages.
In particular, \textsf{R} provides a rich set of built-in as well as extended functions for data extraction, data cleaning, data loading, data transformation, statistical analysis, machine learning and visualization.
In addition, it provides the ability to connect with other languages and systems (e.g., \textsf{Python}).

A main drawback with \textsf{R} is that most of its packages were developed primarily for in-memory and interactive usage, i.e., for scenarios in which the data fit in memory. With the aim of tackling this challenge and providing the ability to handle massive datasets, several systems have been developed to support the execution of R programs on top of the distributed and scalable big data processing platforms such as \textsf{Hadoop} (e.g., \textsf{Ricardo}~\cite{das2010ricardo}, \textsf{RHadoop}\footnote{\url{https://github.com/RevolutionAnalytics/RHadoop}} and \textsf{RHIPE}\footnote{\url{https://github.com/tesseradata/RHIPE}}, \textsf{Segue}\footnote{\url{https://code.google.com/archive/p/segue/}}) and \textsf{Spark}~\cite{SPARK} (e.g., \textsf{SparkR}~\cite{SparkR}). For example,
\textsf{RHIPE} is an R package that brings MapReduce framework to R users and enable them to access the Hadoop cluster from within R environment. In particular, using specific R functions, users become able to launch MapReduce jobs on the Hadoop cluster where the results can be easily retrieved from HDFS.
\textsf{Segue} enables users to execute MapReduce jobs from within R environment on  Amazon Elastic MapReduce platforms.
\textsf{SparkR} has become a popular \textsf{R} package that supports a light-weight frontend to execute  \textsf{R} programs on top of the \textsf{Apache Spark}~\cite{SPARK} distributed computation engine and allows executing large scale data analysis tasks from the \textsf{R} shell.
\textsf{Pydoop}~\cite{leo2010pydoop} is a Python package that provides an API for both the Hadoop framework and the HDFS.
\textsf{Torch7}~\cite{collobert2011torch7} has been presented as a mathematical environment and versatile numeric computing framework for building machine learning algorithms. \textsf{Theano}~\cite{bastien2012theano} has been presented as a linear algebra compiler that optimizes
mathematical computations and produces efficient low-level implementations.

\textsf{SciDB}~\cite{brown2010overview} has been presented as an analytical database which is oriented toward the data management needs of scientific workflows. In particular, it mixes statistical and linear algebra operations with data management operations using a  multi-dimensional array data model.
SciDB supports both a functional (AFL) and a SQL-like query language (AQL) where AQL is compiled into AFL. \textsf{MADlib}~\cite{hellerstein2012madlib} provided a suite of SQL-based implementation for data mining and machine learning algorithms  that are designed to get installed and run at scale within any relational database engine that support extensible SQL, with no need for data import/export to other external tools. \textsf{MLog}~\cite{li2017mlog} has been presented as a high-level language that integrates machine
learning into data management systems.  It extends the query language over the SciDB data model~\cite{brown2010overview} to allow users to specify machine learning models in a way similar to traditional relational views and relational queries. It is designed to manage all data movement, data persistence, and machine-learning related optimizations automatically.

\textsf{H2O}\footnote{http://www.h2o.ai} is an open source framework that provides a parallel processing engine which is equipped with
math and machine learning libraries. It offers support for various programming languages including Java, R, Python, and Scala. \textsf{Apache Mahout}~\cite{owen2012mahout} is an open-source toolkit which is designed  to solve very practical and scalable machine learning problems on top of the \textsf{Hadoop} platform.
Thus, \textsf{Mahout} is primarily meant for distributed and batch processing of massive sizes of data on a cluster.  In particular, Mahout is essentially a set of Java libraries which is well integrated with Apache Hadoop and is designed to make machine learning applications easier to build.
Recently, Mahout has been extended to provide support for machine learning algorithms for collaborative filtering and classification on top of Spark and H2O  platforms. \textsf{MLib}~\cite{meng2016mllib} has been presented as the Spark's~\cite{SPARK} distributed machine learning library  that is well-suited for iterative machine learning tasks. It provides scalable implementations of standard learning algorithms for common
learning settings including classification, regression, collaborative filtering, clustering, and
dimensionality reduction. MLlib supports several languages (e.g., Java, Scala, and Python) and provides a high-level API that leverages Spark's rich ecosystem to simplify the development of end-to-end machine learning pipelines.

Several declarative machine learning implementations have been implemented on top of big data processing systems~\cite{boehm2016declarative}. For example,  \textsf{Samsara}~\cite{schelter2016samsara}, has been introduced as a mathematical environment that supports declarative implementation for general linear algebra and statistical operations as part of the Apache Mahout library.
It  allows its users to specify programs in R-like style using a set of common matrix abstractions and linear algebraic operations.
Samsara compiles, optimizes and executes its programs on distributed dataflow systems (e.g., Apache Spark , Apache Flink, H2O).
\textsf{MLbase}~\cite{kraska2013mlbase} has been implemented to provides a general-purpose machine learning library with a similar goal to Mahout's goal which is to provide a viable solution for dealing with large-scale machine learning tasks on top of  Spark framework.
It supports  a Pig Latin-like~\cite{PigLatin1} declarative language to specify machine learning tasks and implements and provides a set of high-level operators that enable implementing a wide range of machine learning methods without deep systems knowledge. In addition, it implements an optimizer to select and dynamically adapt the choice of learning algorithm. \textsf{Apache SystemML}~\cite{SystemML} provides declarative machine learning framework which is developed to run on top of Apache Spark. It supports R and Python-like syntax that  includes  statistical functions, linear algebra primitives and ML-specific constructs. It  applies cost-based compilation techniques to generate efficient, low-level execution plans with in-memory single-node and large-scale
distributed operations.
\textsf{ScalOps}~\cite{borkar2012declarative} has been presented as a domain-specific language (DSL) that moves beyond single pass data analytics (i.e., MapReduce) to include multi-pass workloads, supporting iteration over algorithms expressed as relational queries on the training and model data.
The physical execution plans of ScalOps consists of dataflow operators which are executed using the Hyracks data-intensive computing engine~\cite{borkar2011hyracks}.
\textsf{Mxnet}~\cite{chen2015mxnet} is a library that has been designed to ease the development
of machine learning algorithms. It blends declarative symbolic expression with imperative tensor computation and offers auto differentiation to derive gradients. MXNet is designed to run on various heterogeneous systems, ranging from mobile devices to distributed GPU clusters.

Microsoft introduced \textsf{AzureML}~\cite{team2016azureml} as a machine learning framework as a software-as-a-service (SaaS) solution which provides a cloud-based visual environment for constructing data analytics workflows.
It is provided as a fully managed service by Microsoft where users neither need to buy any hardware/software nor manually manage any virtual machines.
AzureML provides data scientists with a Web-based machine learning IDE for creating and automating machine learning workflows. In addition, it provides scalable and parallel implementations of popular machine learning techniques as well as data processing capabilities using a drag-and-drop interface. AzureML can read and import  data from various sources including HTTP URL, Azure Blob Storage, Azure Table and Azure SQL Database. It also allows data scientists to import their own custom data analysis scripts (e.g., in R or Python). \textsf{Cumulon}~\cite{huang2013cumulon} has been present as a system which is designed to help users rapidly develop and  deploy matrix-based big-data analysis programs in the cloud. It provides an abstraction for distributed storage of matrices on top of HDFS. In particular, matrices are stored and accessed by tiles.  A Cumulon program executes as a workflow of jobs. Each job reads a number of input matrices and writes a number of output matrices; input and output matrices must be disjoint. Dependencies among jobs are implied by dependent accesses to the same matrices. Dependent jobs execute in serial order. Each job executes as multiple independent tasks that do not communicate with each other. Hadoop-based Cumulon inherits important features of Hadoop such as failure handling, and is able to leverage the vibrant Hadoop ecosystem. While targeting matrix operations, Cumulon can support programs that also contain traditional, non-matrix Hadoop jobs.

\begin{figure*}
  \centering
  \includegraphics[width=1\textwidth]{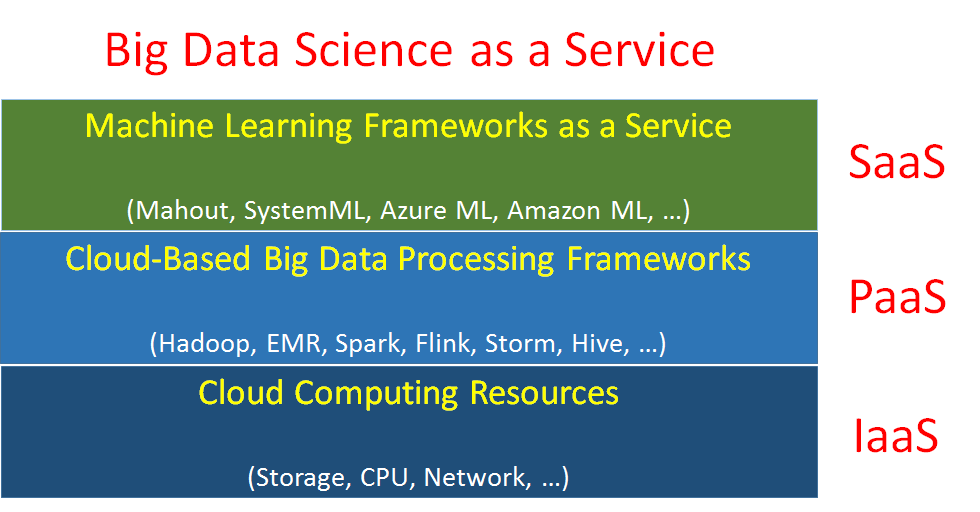}\\
  \caption{Big Data Science as a Service Software Stack}\label{FIG:Stack}
\end{figure*}

Google has also provided a cloud-based SaaS machine learning platform\footnote{\url{https://cloud.google.com/products/machine-learning/}} which is equipped with pre-trained models in addition to a platform to generate users' models. The service is integrated with other Google services such as Google Cloud Storage and Google Cloud Dataflow. It encapsulates powerful machine learning models that supports different analytics applications (e.g. image analysis, speech recognition, text analysis and automatic translation) through  REST API calls.
Similarly, Amazon provide its machine learning as a service solution\footnote{\url{https://aws.amazon.com/machine-learning/}} (AML) which guides its users through the process of creating data analytics  models without the need to learn complex algorithms or technologies. Once the models are created, the service makes it easy to perform predictions via simple APIs without the need to write any user code or manage any hardware or software infrastructure.
AML works with data stored in Amazon S3, RDS or Redshift. It also provides an API set for connecting with and manipulating other data sources.
\textsf{IBM Watson Analytics}\footnote{\url{https://www.ibm.com/analytics/watson-analytics/}} is another SaaS predictive analytic framework that allows its user to express their analytics job using natural English language. The service attempts to automatically spot interesting correlations and exceptions within the input data. It also provides suggestions on the various data cleaning steps and the most adequate data visualization technique to use for various analysis scenarios.

The \textsf{BigML}\footnote{https://bigml.com} SaaS framework supports discovering predictive models from the input data using data classification and regression algorithms. In BigML,  predictive models are presented to the users as an interactive decision tree which is dynamically visualized and explored within the BigML interface. BigML also provides a PaaS solution, \textsf{BigML PredictServer}\footnote{\url{https://bigml.com/predictserver}}, which can be integrated with applications, services, and other data analysis tools. Other SaaS framework include \textsf{Kognitio}\footnote{\url{www.kognitio.com}} and \textsf{Hunk}\footnote{\url{http://www.splunk.com/en_us/products/hunk.html}}. In general, one of the main advantage of working with cloud-based SaaS tools is that users do not have to worry about scaling their solution. Instead, ideally, the provided service should be able to automatically scale if the consumption of computing resources for the defined analytical models  has increased and according to the user defined configurations and requirements.

In practice, the machine learning process involves building complex and multi-stage pipelines that include feature extraction, dimensionality reduction,
data transformations and training supervised learning models. \textsf{Keystoneml} framework~\cite{sparks2017keystoneml} is designed to tackle this challenge by providing a high-level, type-safe API that is built around logical operators to capture end-to-end machine learning applications. To optimize the machine learning pipelines, Keystoneml applies  techniques to do both per-operator optimization and end-to-end pipeline optimization. It uses a cost-based optimizer that accounts for both computation and communication costs. The optimizer is also able  to determine which intermediate states should be materialized
in the main memory during the iterative execution over the raw data. \textsf{Tensorflow}~\cite{abadi2016tensorflow} provides an interface for designing machine learning algorithms, and an implementation for executing such algorithms. In particular, Tensorflow takes computations described using a dataflow-like model and enables compiling them onto several hardware
platforms, ranging from running inference on mobile device platforms (e.g., Android and iOS) to large-scale distributed systems of hundreds
of machines and thousands of computational devices such as
GPU cards. The main focus of Tensorflow is to  simplify the real-world use of machine learning
system and significantly reducing the maintenance burdens. The \textsf{F2} analytics framework~\cite{grandl2017fast} has been designed to separate execution from data management and handles compute and data as equal first-class citizens. In particular, in this framework, data is managed separately while decisions to determine how data is partitioned or when it is to be processed are taken at runtime. The computation that processes the data can have lose semantics
and run any of the available operations on whatever data is ready. One of the main advantages of this framework design is that it provides more flexibility
in expressing analytics jobs by removing concerns regarding data partitioning, routing and what logic to specify during the runtime.

\section{Discussion and Open Challenges}

The world is progressively moving towards being a data-driven society where data are the most valuable asset.   The proliferation of big data and big computing boosted the adoption of machine learning and data science across several application domains. In practice, efficient and effective analysis and exploitation of Big Data have become essential requirements for enhancing the competitiveness of enterprises and maintaining sustained
social and economic growth of societies and countries.
Therefore, Big Data Science has become a very active research domain with crucial impact on various scientific and business domains where it is significant to analyze massive and complex amounts of data. In practice, in many cases data to be analyzed can be stored in cloud services and elastic computing cloud resources can be exploited to facilitate the speeding up and scaling out of the data science tasks.

Figure~\ref{FIG:Stack} illustrates the building blocks of the Big Data Science as a Service software stack.
In spite of the high expectations on the promises and potential benefits of Big Data science, there are still many challenges to overcome to be able to fully harness its full power. For example, in practice, big data science lives and dies by the data. It mainly rests on the availability of massive datasets, of that there can be no doubt. The more data that is available, the richer the insights and the results that big data science can produce. The bigger and more diverse the data set, the better the analysis can model the real world. Therefore, any successful big data science process has attempted to incorporate as many data sets from internal and public sources as possible. In reality, data is segmented, siloed and under the control of different individuals, departments or organizations. It is  crucially required to motivate all parties to work collaboratively and share useful data/insights for the public. Recently, there has been an increasing trend for open data initiatives which supports the idea of making data publicly available to everyone to use and republish as they wish, without restrictions from copyright, patents or other mechanisms of control~\cite{huijboom2011open}.
Online data markets~\cite{BalazinskaHS11} are emerging cloud-based services (e.g., \textsf{Azure Data Market}\footnote{\url{http://datamarket.azure.com/browse/data}}, \textsf{Kaggle}\footnote{\url{https://www.kaggle.com/}}, \textsf{Connect}\footnote{\url{https://connect.data.com/}}, \textsf{Socrata}\footnote{\url{https://socrata.com/}}).
For example, Kaggle is a platform where companies can provide data to a community of data scientists so that they can analyze the data with the aim of discovering predictive, actionable insights and win incentive awards. In particular, such platforms follow a model where data and rewards are traded for innovation. More research, effort and development is still required in this direction.

With the increasing number of platforms and services, interoperability is arising as a main issue. Standard formats and models are required to enable interoperability and ease cooperation among the various platforms and services.  In addition, the service-oriented paradigm can play an effective roles in supporting the execution of large-scale distributed analytics on heterogeneous platforms along with software components developed using various programming languages or tools. Furthermore, in practice, the majority of existing big-data-processing platforms (e.g., Hadoop and Spark) are designed based on the single-cluster setup with the assumptions of centralized management and homogeneous connectivity which makes them sub-optimal and sometimes infeasible to apply for scenarios that require implementing data analytics jobs on highly distributed data sets (e.g., across racks, clusters, data centers or multi-organizations).
Some scenarios can also require distributing data analysis tasks in a hybrid mode among local processing of local data sources and model exchange and fusion mechanisms to compose the results produced in the distributed nodes.

In practice, the design of most of the statistical computation (e.g., R) and scientific computing tools (e.g. Python) is memory-bounded where data analysis algorithms rely on the in-memory data processing mechanism. While this approach may bring many benefits in terms of processing speed up and faster decisions, with big data sizes, there could be scalability risks due to performance issues if the processed data do not fit in the available main memory or it can be very costly if the required memory can be allocated, in a cloud environment. Efficient distributed and parallel disk-based execution mechanisms for complex data analysis jobs are crucially required to tackle this challenge.

In general, a major obstacle for supporting big data analytics applications is the challenging and time consuming process of identifying and training an adequate predictive model. Therefore, data science is a highly iterative exploratory process where most scientists work hard to find the best model or algorithms that meets their data challenge. In practice, there is no one-model-fits-all solution, thus, there is no single model or algorithm that can handle all data set varieties and changes in data that may occur over time. All machine learning algorithms require user defined inputs to achieve a balance between accuracy and generalizability, referred to as tuning
parameters. These tuning parameters impact the way the algorithm searches
for the optimal solution. Recent research efforts have been attempting to automate this process. However, they have mainly focused on single node
implementations and have assumed that model training itself is a black box, limiting their usefulness for applications driven by large-scale datasets~\cite{KumarMNP15}.  Recently, the \textsf{ModelHub}~\cite{miao2016modelhub} has been proposed to tackle parts of this problem. In particular, it provides a a model versioning system (DLV) to store and query the models and their versions, a domain specific language that serves as an abstraction layer for searching through model space  in addition to a hosted service to store developed models, explore existing models, enumerate new models and share models with others. \textsf{ModelDB}~\cite{vartak2016m} is another system for managing machine learning models that automatically tracks the models in their native
environments (e.g. Mahout, SparkML), indexes them  and allows flexible exploration of models using either SQL or a visual web-based interface.
Along with models and pipelines, ModelDB  stores several metadata (e.g., parameters of pre-processing steps, hyperparameters for models etc.) and quality metrics (e.g. AUC, accuracy). In addition, it can store the training and test data for each model.
The \textsf{DAWN} project at Stanford~\cite{bailis2017infrastructure} has recently announced its vision for the next five years with the aim of making  the machine learning (ML) process \emph{usable} for small teams of non-ML experts so that they can easily apply ML to their problems, achieve high-quality results and deploy production systems that can be used in critical applications. The main design philosophy of the DAWN project is to target the management of end-to-end ML workflows, empower domain experts to easily develop and deploy their models and perform effective optimization of the workflow execution pipelines. We believe that additional research efforts are crucially required to efficiently and effectively manage the life cycle and pipelines of the data science process.
\bibliographystyle{abbrv}

\end{document}